\documentclass[12pt]{article}

\usepackage{graphicx}
\usepackage{float}
\usepackage{cite}

\def\gtwid{\mathrel{\raise.3ex\hbox{$>$\kern-.75em\lower1ex\hbox{$\sim$}}}}
\def\ltwid{\mathrel{\raise.3ex\hbox{$<$\kern-.75em\lower1ex\hbox{$\sim$}}}}
\def\square{\kern1pt\vbox{\hrule height 1.2pt\hbox{\vrule width 1.2pt\hskip 3pt
   \vbox{\vskip 6pt}\hskip 3pt\vrule width 0.6pt}\hrule height 0.6pt}\kern1pt}

\begin{document}

\begin{titlepage}

\begin{flushright}
UFIFT-QG-19-03 \\
CP3-19-39
\end{flushright}

\vskip 1cm

\begin{center}
{\bf Graviton Propagator in a 2-Parameter Family of de Sitter Breaking Gauges}
\end{center}

\vskip .5cm

\begin{center}
D. Glavan$^{1*}$, S. P. Miao$^{2\star}$, T. Prokopec$^{3\dagger}$ and
R. P. Woodard$^{4\ddagger}$
\end{center}

\vskip .5cm

\begin{center}
\it{$^{1}$ Centre for Cosmology, Particle Physics and Phenomenology (CP3) \\
Universit\'e catholique de Louvain, Chemin du Cyclotron 2, 1348 Louvain-la-Neuve, BELGIUM}
\end{center}

\begin{center}
\it{$^{2}$ Department of Physics, National Cheng Kung University \\
No. 1, University Road, Tainan City 70101, TAIWAN}
\end{center}

\begin{center}
\it{$^{3}$ Institute for Theoretical Physics, Spinoza Institute \& EMME$\Phi$ \\
Utrecht University, Postbus 80.195, 3508 TD Utrecht, THE NETHERLANDS}
\end{center}

\begin{center}
\it{$^{4}$ Department of Physics, University of Florida,\\
Gainesville, FL 32611, UNITED STATES}
\end{center}

\vspace{.5cm}

\begin{center}
ABSTRACT
\end{center}

We formulate the graviton propagator on de Sitter background in a 2-parameter
family of simple gauges which break de Sitter invariance. Explicit results are
derived for the first order perturbations in each parameter. These results
should be useful in computations to check for gauge dependence of graviton
loop corrections.

\begin{flushleft}
PACS numbers: 04.50.Kd, 95.35.+d, 98.62.-g
\end{flushleft}

\vskip .5cm

\begin{flushleft}
$^{*}$ e-mail: drazen.glavan@uclouvain.be \\
$^{\star}$ e-mail: spmiao5@mail.ncku.edu.tw \\
$^{\dagger}$ e-mail: T.Prokopec@uu.nl \\
$^{\ddagger}$ e-mail: woodard@phys.ufl.edu
\end{flushleft}

\end{titlepage}

\section{Introduction} \label{intro}

Explicit computations of 1PI (one-particle-irreducible) 2-point functions
on de Sitter background \cite{Tsamis:1996qk,Miao:2005am,Kahya:2007bc,
Miao:2012bj,Leonard:2013xsa}, and their use to quantum-correct the linearized
effective field equations, provide strong evidence that inflationary 
gravitons modify particle kinematics \cite{Mora:2013ypa,Miao:2006gj,
Kahya:2007cm,Wang:2014tza} and change force laws \cite{Glavan:2013jca}.
However, there is a persistent anxiety about the reality of these effects
owing to the possibility of dependence on the gauge used to define the
graviton propagator. The computations mentioned above were all made in 
a very simple gauge that breaks de Sitter invariance \cite{Tsamis:1992xa,
Woodard:2004ut}. When the vacuum polarization was computed in a vastly
more complicated, 1-parameter family of de Sitter invariant gauges
\cite{Mora:2012zi}, the result looks very different \cite{Glavan:2015ura},
and the enhancement it provides to dynamical photons is slightly different
\cite{Glavan:2016bvp}, although it has the same sign and time dependence.

We seek to establish the reality of graviton corrections to particle kinematics
and force laws by purging the linearized effective field equations of gauge 
dependence. We are developing a technique for accomplishing this by including 
generic parts of the quantum gravitational correlations with the source that
disturbs the effective field and with the observer who measures the disturbance
\cite{Miao:2017feh}. The technique relies on the position-space version of a 
set of identities derived by Donoghue \cite{Donoghue:1993eb,Donoghue:1994dn,
Donoghue:1996mt} that allow one to view the infrared singular parts of an 
invariant amplitude as corrections to the 1PI 2-point function. We have already 
shown that the technique works, for graviton corrections to massless scalar
exchange on flat space background, by making the computation in the 2-parameter
family of Poincar\'e invariant gauges and demonstrating that the corrected 1PI
2-point function is independent of the gauge parameters \cite{Miao:2017feh}.
Our goal is to carry out the same computation on de Sitter background, for which
we require a generalization of the 2-parameter family of flat space gauges. 
Providing that generalization is the point of this paper.

In section~\ref{2pgauge} we review the simple gauge and propose an appropriate 
2-parameter generalization. Constructing the graviton propagator as a general 
function of these two parameters is challenging, but almost as much information 
about gauge dependence can be gained by deriving just the first order variations 
about the simple gauge. That is done in section~\ref{solution}. We discuss the 
results in section~\ref{discuss}. Section~\ref{append} consists of an appendix
in which we derive explicit forms for the various integrated propagators required
for the solution.

\section{A 2-Parameter Family of Gauges} \label{2pgauge}

Our de Sitter background geometry is given in conformal coordinates on $D$ spacetime
dimensions with spacelike signature,
\begin{equation}
ds^2 = a^2 \Bigl[-d\eta^2 + d\vec{x} \!\cdot\! d\vec{x}\Bigr] \qquad , \qquad
a(\eta) = -\frac{1}{H \eta} \;\; (\eta < 0) \; . \label{deSitter}
\end{equation}
The graviton field comes from the conformally rescaled metric, $g_{\mu\nu}(x) 
\equiv a^2 [ \eta_{\mu\nu} + \kappa h_{\mu\nu}(x)]$. Here $\kappa^2 \equiv 16 
\pi G$ and graviton indices are raised and lowered using the Minkowski metric, 
$h^{\mu}_{~\nu} \equiv \eta^{\mu\rho} h_{\rho\nu}$. 

The simple propagator is defined by the gauge fixing term \cite{Tsamis:1992xa,
Woodard:2004ut},
\begin{equation}
\mathcal{L}_{\rm GF} = -\frac12 a^{D-2} \eta^{\mu\nu} F_{\mu} F_{\nu} \quad , 
\quad F_{\mu} = \eta^{\rho\sigma} \Bigl( h_{\mu\rho , \sigma} - \frac12 
h_{\rho\sigma ,\mu} + (D \!-\! 2) H a h_{\mu\rho} \delta^0_{~\sigma}\Bigr) .
\label{simplegauge}
\end{equation}
The graviton propagator in this gauge takes the form,
\begin{equation}
i\Bigl[\mbox{}_{\mu\nu} \Delta_{\rho\sigma}\Bigr](x;x') = \sum_{I=A,B,C} 
\Bigl[ \mbox{}_{\mu\nu} T^I_{\rho\sigma}\Bigr] \times i \Delta_I(x;x') \; .
\label{simpleprop}
\end{equation}
The three propagators $i\Delta_I(x;x')$ are for minimally coupled scalars with
masses $M^2_A = 0$, $M^2_B = (D-2) H^2$ and $M^2_C = 2(D-3) H^2$. The three 
tensor factors are constructed from $\eta_{\mu\nu}$, $\overline{\eta}_{\mu\nu} 
\equiv \eta_{\mu\nu} + \delta^0_{~\mu} \delta^0_{~\nu}$ and $\delta^0_{~\mu}$,
\begin{eqnarray}
\Bigl[\mbox{}_{\mu\nu} T^A_{\rho\sigma}\Bigr] & = & 2 \, \overline{\eta}_{\mu (\rho} 
\overline{\eta}_{\sigma) \nu} - \frac{2}{D \!-\! 3} \, \overline{\eta}_{\mu\nu}
\overline{\eta}_{\rho\sigma} \quad , \quad \Bigl[\mbox{}_{\mu\nu} T^B_{\rho\sigma}
\Bigr] = -4 \delta^0_{~(\mu} \overline{\eta}_{\nu ) (\rho} \delta^0_{~\sigma)} 
\; , \label{TAB} \\
\Bigl[\mbox{}_{\mu\nu} T^C_{\rho\sigma}\Bigr] & = & \frac{2 E_{\mu\nu}
E_{\rho\sigma}}{(D \!-\! 2) (D \!-\! 3)} \quad , \quad E_{\mu\nu} \equiv 
(D \!-\!3) \delta^0_{~\mu} \delta^0_{~\nu} \!+\! \overline{\eta}_{\mu\nu} 
\; , \label{TC}  
\end{eqnarray}
where parenthesized indices are symmetrized. The simple graviton propagator
(\ref{simpleprop}) is easy to use for three reasons:
\begin{enumerate}
\item{In $D = 4$ its three propagators consist of just one or two terms
involving the two scale factors and the invariant interval of flat space,
\begin{equation}
i\Delta_A \longrightarrow \frac1{4\pi^2} \Biggl[ \frac1{a a' \Delta x^2}
-\frac{H^2}{2} \ln(H^2 \Delta x^2) \Biggr] \; , \; i\Delta_{B,C} \longrightarrow 
\frac1{4 \pi^2 a a' \Delta x^2} \; ; \label{DABC4D}
\end{equation}}
\item{Its tensor factors are constants; and}
\item{Its 1PI 2-point functions are elementary functions of $\Delta x^2$, $a$ 
and $a'$.}
\end{enumerate}
\noindent None of these features pertains for de Sitter invariant gauges 
\cite{Mora:2012zi,Glavan:2015ura}, which is why only a single loop computation has 
been made using them.

It is desirable to construct the graviton propagator in a 2-parameter 
family of de Sitter breaking gauges that can be seen as perturbations of the 
simple gauge \cite{Tsamis:1992xa,Woodard:2004ut}. This also permits a 2-parameter
family of gauges rather than the 1-parameter family that would be available 
without de Sitter breaking, owing to a topological obstacle which precludes adding 
de Sitter invariant gauge fixing functionals \cite{Miao:2009hb}. Finally, it is 
advantageous that the flat space limit of our gauge condition should agree with the 
2-parameter family of gauges used for the flat space computation \cite{Miao:2017feh} 
which we seek to generalize to de Sitter. A plausible generalization of 
(\ref{simplegauge}) is therefore,
\begin{equation}
\mathcal{L}^{\alpha\beta}_{\rm GF} = -\frac{a^{D-2}}{2\alpha} \eta^{\mu\nu} 
\mathcal{F}_{\mu} \mathcal{F}_{\nu} \; , \; \mathcal{F}_{\mu} = 
\eta^{\rho\sigma} \Bigl( h_{\mu\rho , \sigma} - \frac{\beta}{2} 
h_{\rho\sigma ,\mu} + (D - 2) H a h_{\mu\rho} \delta^0_{~\sigma}\Bigr) . 
\label{newgauge}
\end{equation}
Note that taking $\alpha = \beta = 1$ corresponds to the simple gauge 
(\ref{simplegauge}), and that taking the flat space limit ($H=0$ and $a = 1$)
recovers the family of gauges in which the flat space calculation 
\cite{Miao:2017feh} was made.

\section{Our Solution for the Propagator} \label{solution}

The purpose of this section is to construct the first order perturbations 
in $\delta \alpha \equiv \alpha - 1$ and $\delta \beta \equiv \beta - 1$ 
of the propagator in the gauge (\ref{newgauge}). We first review the flat
space propagator, and its use in studies of gauge dependence \cite{Miao:2017feh},
to make two points:
\begin{enumerate}
\item{The result will involve convolutions of propagators; and}
\item{First order perturbations in the gauge parameters provide almost as 
much information about gauge as the all-orders result.}
\end{enumerate}
\noindent We then return to de Sitter to define the necessary integrated 
propagators, whose evaluation is consigned to the Appendix. The section closes
with results for the first order perturbations in $\delta \alpha$ and $\delta
\beta$. 

\subsection{Lessons from Flat Space} \label{flat}

In the flat space limit of (\ref{newgauge}) the graviton propagator is
\cite{Capper:1979ej},
\begin{eqnarray}
\lefteqn{i\Bigl[\mbox{}_{\mu\nu} \Delta^{\rm flat}_{\rho\sigma}\Bigr](x;x') 
= \Biggl\{ 2 \Pi_{\mu (\rho} \Pi_{\sigma) \nu} \!-\! \frac2{D \!-\! 1} 
\Pi_{\mu\nu} \Pi_{\rho\sigma} } \nonumber \\
& & \hspace{.2cm} -\frac2{(D \!-\! 2)(D \!-\! 1)} \Biggl[ \eta_{\mu\nu}
\!-\! \Bigl( \frac{D \beta \!-\! 2}{\beta \!-\! 2} \Bigr) \frac{\partial_{\mu}
\partial_{\nu} }{\partial^2} \Biggr] \Biggl[ \eta_{\rho\sigma}
\!-\! \Bigl( \frac{D \beta \!-\! 2}{\beta \!-\! 2} \Bigr) \frac{\partial_{\rho}
\partial_{\sigma} }{\partial^2} \Biggr] \nonumber \\
& & \hspace{2.5cm} + 4\alpha \times \frac{\partial_{(\mu} \Pi_{\nu) (\rho}
\partial_{\sigma)} }{\partial^2} + \frac{4 \alpha}{(\beta \!-\! 2)^2} \times
\frac{\partial_{\mu} \partial_{\nu} \partial_{\rho} \partial_{\sigma}}{
\partial^4} \Biggr\} i\Delta(x;x') \; . \qquad \label{flatgrav}
\end{eqnarray}
Here the transverse projection operator is $\Pi_{\mu\nu} \equiv \eta_{\mu\nu} 
- \frac{\partial_{\mu} \partial_{\nu}}{\partial^2}$ and the massless scalar
propagator in flat space is,
\begin{equation}
i\Delta(x;x') = \frac{\Gamma(\frac{D}2 \!-\! 1)}{4 \pi^{\frac{D}{2}} \Delta x^{D-2}}
\qquad \Longrightarrow \qquad \partial^2 i\Delta(x;x') = i \delta^D(x \!-\! x') 
\; . \label{flatDelta}
\end{equation}
First note from the inverse powers of $\partial^2$ which act on $i\Delta(x;x')$
that the general gauge propagator (\ref{flatgrav}) involves the convolution of
one propagator with another \cite{Leonard:2012fs},\footnote{The $\Delta x$-dependent
part of (\ref{flatDelta}) follows from the usual procedure of combining denominators,
shifting, Wick-rotating and evaluating the Euclidean-space integration,
\begin{eqnarray}
\lefteqn{-i \!\! \int \!\! d^Dz \, \frac{\Gamma(\frac{D}2 \!-\! 1)}{4 \pi^{\frac{D}2}}
\Bigl[ \frac1{(x \!-\! z)^2 \!+\! i \epsilon}\Bigr]^{\frac{D}2-1} \times
\frac{\Gamma(\frac{D}2 \!-\! 1)}{4 \pi^{\frac{D}2}} \Bigl[ \frac1{(z \!-\! x')^2
\!+\! i \epsilon}\Bigr]^{\frac{D}2 -1} } \nonumber \\
& & = -\frac{i \Gamma(D \!-\!2)}{16 \pi^{D}} \int_{0}^{1} \!\! ds \, s^{\frac{D}2 -2} 
(1 \!-\! s)^{\frac{D}2 - 2} \int \!\! \frac{d^Dz}{ [s (x \!-\! z)^2 + (1 \!-\! s) 
(z \!-\! x')^2 + i \epsilon]^{D-2}} \nonumber \\
& & = -\frac{\Gamma(D \!-\! 2)}{16 \pi^D} \int_0^1 \!\! ds \, s^{\frac{D}2 -2}
(1 \!-\! s)^{\frac{D}2 -2} \int \!\! \frac{d^Dz_E}{[z^2_E \!+\! s (1 \!-\! s)
\Delta x^2]^{D-2}} \; , \nonumber \\
& & = -\frac{\Gamma(D \!-\! 2)}{16 \pi^D} \int_0^1 \!\! ds \, s^{\frac{D}2 -2}
(1 \!-\! s)^{\frac{D}2 -2} \times \frac{\pi^{\frac{D}2}}{\Gamma(\frac{D}2) [s (1 \!-\! s) 
\Delta x^2]^{\frac{D}2 -2}} \times \frac{\Gamma(\frac{D}2 \!-\! 2) \Gamma(\frac{D}2)}{
\Gamma(D \!-\! 2)} \; . \nonumber
\end{eqnarray}}
\begin{equation}
\frac1{\partial^2} i\Delta(x;x') = -i \!\! \int \!\! d^Dz \, i\Delta(x;z) 
i\Delta(z;x') = \frac{\Gamma(\frac{D}2 \!-\! 2)}{16 \pi^{\frac{D}{2}}} \Bigl[
\mu^{D-4} - \frac1{\Delta x^{D-4}}\Bigr] , \label{flatintprop}
\end{equation}
where $\mu$ is a regularization scale.
The flat space limit therefore implies that similar convolutions must occur in 
the general gauge propagator on de Sitter background.

Now expand the gauge parameters around their ``simple'' values,
\begin{equation}
\alpha \equiv 1 + \delta \alpha \qquad , \qquad \beta \equiv 1 + \delta \beta \; ,
\end{equation}
Expanding the general gauge propagator (\ref{flatgrav}) in $\delta \alpha$ and
$\delta \beta$ gives,
\begin{eqnarray}
\lefteqn{i\Bigl[\mbox{}_{\mu\nu} \Delta^{\rm flat}_{\rho\sigma}\Bigr] =
\Biggl[2 \eta_{\mu (\rho} \eta_{\sigma) \nu} - \frac{2 \eta_{\mu\nu} 
\eta_{\rho\sigma}}{D \!-\! 2}  \Biggr] i\Delta(x;x') + \delta \alpha \!\times\! 
\Biggl[ \frac{4 \partial_{(\mu} \eta_{\nu) (\rho} \partial_{\sigma)} }{\partial^2} 
\Biggr] i\Delta(x;x') } \nonumber \\
& & \hspace{1.3cm} -\delta \beta \!\times\! \frac{4}{D \!-\! 2} \Biggl[ 
\frac{\eta_{\mu\nu} \partial_{\rho} \partial_{\sigma}}{\partial^2} + 
\frac{\partial_{\mu} \partial_{\nu} \eta_{\rho\sigma}}{\partial^2} \Biggr] 
i\Delta(x;x') + O\Bigl(\delta \alpha \delta \beta,\delta \beta^2\Bigr) \; . 
\qquad \label{flatalphabeta}
\end{eqnarray}
Table~\ref{Cab} concerns gauge dependence in one graviton loop corrections to the 
effective field equation for a massless, minimally coupled scalar on flat space
background \cite{Miao:2017feh}.
\begin{table}[ht]
\setlength{\tabcolsep}{8pt}
\def\arraystretch{1.5}
\begin{minipage}{0.6\textwidth}
\centering
\begin{tabular}{|@{\hskip 1mm }c@{\hskip 1mm }||c|c|c|c|}
\hline
$i$ & $1$ & $\alpha$ & $\frac1{\beta -2}$ & $\frac{(\alpha-3)}{(\beta-2)^2}$ \\
\hline\hline
0 & $+\frac34$ & $-\frac34$ & $-\frac32$ & $+\frac34$ \\
\hline
1 & $0$ & $0$ & $0$ & $+1$ \\
\hline
2 & $0$ & $0$ & $0$ & $0$ \\
\hline
3 & $0$ & $0$ & $+3$ & $-2$ \\
\hline
4 & $+\frac{17}4$ & $-\frac34$ & $0$ & $-\frac14$ \\
\hline
5 & $-2$ & $+\frac32$ & $-\frac32$ & $+\frac12$ \\
\hline\hline
Total & $+3$ & $0$ & $0$ & $0$ \\
\hline
\end{tabular}
\end{minipage}
\begin{minipage}{0.6\textwidth}
\begin{tabular}{|@{\hskip 1mm }c@{\hskip 1mm }||c|c|c|}
\hline
$i$ & $1$ & $\delta \alpha$ & $\delta \beta$ \\
\hline\hline
0 & $0$ & $0$ & $-\frac32$ \\
\hline
1 & $-2$ & $+1$ & $-4$ \\
\hline
2 & $0$ & $0$ & $0$ \\
\hline
3 & $+1$ & $-2$ & $+5$ \\
\hline
4 & $+4$ & $-1$ & $+1$ \\
\hline
5 & $0$ & $+2$ & $-\frac12$ \\
\hline\hline
Total & $+3$ & $0$ & $0$ \\
\hline
\end{tabular}
\end{minipage}
\caption{\footnotesize The left hand table gives the various gauge-dependent 
multiplicative factors for one graviton corrections to the massless scalar 
exchange force on flat space background using the full propagator (\ref{flatgrav}) 
on the left, and for the first order expansion (\ref{flatalphabeta}) on the right. 
The contribution of the 1PI 2-point function is $i=0$ and the other values of $i$ 
correspond to different source and observer corrections that are precisely 
defined in \cite{Miao:2017feh} and which are necessary to eliminate gauge 
dependence.}
\label{Cab}
\end{table}
The highly gauge dependent contribution from the 1PI 2-point function is $i=0$,
and the five source and observer corrections are reported, both for the full
propagator (\ref{flatgrav}) on the left, and for the first three terms in the 
expansion (\ref{flatalphabeta}) on the right. The left hand table follows the
cancellation of three distinct combinations of the gauge parameters: $\alpha$,
$\frac1{\beta - 2}$ and $\frac{(\alpha - 3)}{(\beta - 2)^2}$. The right hand
table follows two: $\delta \alpha$ and $\delta \beta$. So the vastly simpler 
first order perturbations (\ref{flatalphabeta}) provide two thirds of the checks
available from the full propagator (\ref{flatgrav}).

\subsection{Integrated Propagators in de Sitter} \label{intprops}

The insights we have just derived from the flat space limit motivate constructing
just the order $\delta \alpha$ and $\delta \beta$ perturbations of the 
(\ref{newgauge}) propagator,
\begin{equation}
i\Bigl[\mbox{}_{\mu\nu} \Delta_{\rho\sigma}\Bigr] =
i\Bigl[\mbox{}_{\mu\nu} \Delta^0_{\rho\sigma}\Bigr] +
\delta \alpha \times i\Bigl[\mbox{}_{\mu\nu} \Delta^{\alpha}_{\rho\sigma}\Bigr] +
\delta \beta \times i\Bigl[\mbox{}_{\mu\nu} \Delta^{\beta}_{\rho\sigma}\Bigr] +
O(\delta^2) \; . \label{genform}
\end{equation}
The flat space limit also implies that these perturbations must involve integrated 
propagators analogous to (\ref{flatintprop}). To understand more fully what these are,
recall that the 0-th order propagator $i[\mbox{}_{\mu\nu} \Delta^0_{\rho\sigma}](x;x')$
takes the form (\ref{simpleprop}) where the tensor factors (\ref{TAB}-\ref{TC}) 
represent a sort of $3+1$ decomposition into purely spatial ($A$-type), mixed space 
and time ($B$-type), and temporal plus trace ($C$-type) terms. The corresponding $A$,
$B$ and $C$ scalar propagators are most easily represented in terms of the de Sitter 
length function $y(x;x')$,
\begin{equation}
y(x;x') \equiv H^2 a(\eta) a(\eta') \Biggl[ \Bigl\Vert \vec{x} \!-\! \vec{x}' 
\Bigr\Vert^2 - \Bigl( \vert \eta \!-\! \eta'\vert - i \epsilon\Bigr)^2 \Biggr] 
\; . \label{ydef}
\end{equation}
The $A$-type propagator breaks de Sitter invariance \cite{Onemli:2002hr,Onemli:2004mb},
\begin{eqnarray}
\lefteqn{i \Delta_A \!=\! \frac{H^{D-2}}{(4\pi)^{\frac{D}2}} \Biggl\{
\frac{\Gamma(\frac{D}2)}{\frac{D}2 \!-\! 1} \Bigl( \frac{4}{y}\Bigr)^{\!\frac{D}2-1}
\!\!\!\!\!+\! \frac{\Gamma(\frac{D}2 \!+\! 1)}{\frac{D}2 \!-\! 2} 
\Bigl(\frac{4}{y}\Bigr)^{\!\frac{D}2-2} \!\!\!\!\!\!\!-\! 
\frac{\Gamma(D\!-\!1)}{\Gamma(\frac{D}2)} \Biggl[ \pi \cot\Bigl(\pi 
\frac{D}{2}\Bigr) \!-\! \ln(a a')\Biggr] } \nonumber \\
& & \hspace{1cm} + \sum_{n=1}^{\infty} \Biggl[
\frac1{n} \frac{\Gamma(n \!+\! D \!-\! 1)}{\Gamma(n \!+\! \frac{D}2)}
\Bigl(\frac{y}4 \Bigr)^n \!\!\!\! - \frac1{n \!-\! \frac{D}2 \!+\! 2}
\frac{\Gamma(n \!+\!  \frac{D}2 \!+\! 1)}{\Gamma(n \!+\! 2)} \Bigl(\frac{y}4
\Bigr)^{n - \frac{D}2 +2} \Biggr] \Biggr\} . \qquad \label{Adef}
\end{eqnarray}
In contrast, the $B$-type and $C$-type propagators are de Sitter 
invariant,\footnote{The infinite sums (\ref{Adef}-\ref{Cdef}) might seem 
intimidating but the simple $D=4$ limits (\ref{DABC4D}) mean that only a 
few of the lowest terms need to be retained when divergences are present.} 
\begin{equation}
i \Delta_B = \frac{H^{D-2}}{(4\pi)^{\frac{D}2}} \Biggl\{
\frac{\Gamma(\frac{D}2)}{\frac{D}2 \!-\! 1} \Bigl( \frac{4}{y}\Bigr)^{\frac{D}2-1}
\!\!\!\!\!+\! \sum_{n=0}^{\infty} \Biggl[ \frac{\Gamma(n \!+\! 
\frac{D}2)}{\Gamma(n \!+\!2)} \Bigl( \frac{y}4 \Bigr)^{n - \frac{D}2 +2} 
\!\!\!\!\!-\! \frac{\Gamma(n \!+\! D \!-\! 2)}{\Gamma(n \!+\! \frac{D}2)} 
\Bigl(\frac{y}4 \Bigr)^n \Biggr] \Biggr\} , \label{Bdef}
\end{equation}
\begin{eqnarray}
\lefteqn{i \Delta_C = \frac{H^{D-2}}{(4\pi)^{\frac{D}2}} \Biggl\{ 
\frac{\Gamma(\frac{D}2)}{\frac{D}2 \!-\! 1} \Bigl( \frac{4}{y}\Bigr)^{\frac{D}2-1}
\!\!\!\!- \sum_{n=0}^{\infty} \Biggl[ \Bigl(n \!-\! \frac{D}2 \!+\!  3\Bigr) 
\frac{\Gamma(n \!+\! \frac{D}2 \!-\! 1)}{\Gamma(n \!+\! 2)} \Bigl(\frac{y}4
\Bigr)^{n - \frac{D}2 +2} } \nonumber \\
& & \hspace{6.5cm} - (n\!+\!1) \frac{\Gamma(n \!+\! D \!-\! 3)}{\Gamma(n \!+\! 
\frac{D}2)} \Bigl(\frac{y}4 \Bigr)^n \Biggr] \Biggr\} . \qquad \label{Cdef}
\end{eqnarray}
We obviously require the convolution of any pair of these three propagators.
There is also the matter of derivatives, and the crucial factors of $1/a$ with
which they are associated on de Sitter. Although the integrated propagators of 
flat space (\ref{flatalphabeta}) always carry two derivatives $\frac{\partial_{\mu} 
\partial_{\nu}}{\partial^2} \times i\Delta(x;x')$, those on de Sitter can also have
one derivative or none. Hence our integrated propagators on de Sitter involve a 
measure factor of $a^D$, divided by zero, one, or two powers of $a$,
\begin{eqnarray}
I_{\mu\nu}(x;x') & \equiv & -i\! \int \! d^Dz \, a_z^{D}
i\Delta_{\mu}(x;z) i \Delta_{\nu}(z;x') \; , \label{Idef} \\
J_{\mu\nu}(x;x') & \equiv & -i\! \int \! d^Dz \, a_z^{D-1}
i\Delta_{\mu}(x;z) i \Delta_{\nu}(z;x') \; , \label{Jdef} \\
K_{\mu\nu}(x;x') & \equiv & -i\! \int \! d^Dz \, a_z^{D-2}
i\Delta_{\mu}(x;z) i \Delta_{\nu}(z;x') \; . \label{Kdef}
\end{eqnarray}
The Appendix derives explicit results for these expressions.

As explained, the $I_{\mu\nu}(x;x')$ integrals carry no derivatives, the 
$J_{\mu\nu}(x;x')$ carry one derivative, and the $K_{\mu\nu}(x;x')$ integrals 
carry two derivatives. Although these derivatives could be reflected outside 
the $z^{\mu}$ integration to act on either of the external variables $x^{\mu}$ 
or ${x'}^{\mu}$, the result is cumbersome. (The procedure is explained in the 
Appendix.) It is also possible that the effort would be wasted if reducing 
the diagram in which the perturbed propagator resides would be facilitated by 
retaining the original derivatives, or by reflecting them to the opposite 
coordinate. We have therefore devised a notation in which the symbol ``$D_{\mu}$''
that $\frac{\partial}{\partial z^{\mu}}$ acts on the left hand propagator.
The notation ``$\mathcal{D}_{\mu}$'' indicates that $\frac{\partial}{\partial 
z^{\mu}}$ acts on the right hand propagator. Some examples are,
\begin{eqnarray}
D_{\mu} J_{\rho\sigma}(x;x') & \equiv & -i\! \int \! d^Dz \, a_z^{D-1}
\frac{\partial i\Delta_{\rho}(x;z)}{\partial z^{\mu}} i \Delta_{\sigma}(z;x') 
\; , \label{DJ} \\
\mathcal{D}_{\mu} J_{\rho\sigma}(x;x') & \equiv & -i\! \int \! d^Dz \, a_z^{D-1}
i\Delta_{\rho}(x;z) \frac{\partial i\Delta_{\sigma}(z;x')}{\partial z^{\mu}} 
\; , \label{scriptDJ} \\
D_{\mu} D_{\nu} K_{\rho\sigma}(x;x') & \equiv & -i\! \int \! d^Dz \, a_z^{D-2}
\frac{\partial^2 i\Delta_{\mu}(x;z)}{\partial z^{\mu} \partial z^{\nu}} 
i \Delta_{\nu}(z;x') \; . \label{DDK}
\end{eqnarray}

\subsection{The $\delta \alpha$ and $\delta \beta$ Perturbations} \label{delalphabeta}

Suppose we invert a full kinetic operator $\mathbf{D}$ which can be expressed as
the sum of a 0-th order $\mathbf{D}_0$ operator and a perturbation $\mathbf{D}_1$.
The full propagator can be expanded in familiar geometric series,
\begin{equation}
\frac{i}{\mathbf{D}_0 + \mathbf{D}_1} = \frac{i}{\mathbf{D}_0} + \frac{i}{\mathbf{D}_0}
\times i\mathbf{D}_1 \times \frac{i}{\mathbf{D}_0} + \frac{i}{\mathbf{D}_0} \times 
i\mathbf{D}_1 \times \frac{i}{\mathbf{D}_0} \times i\mathbf{D}_1 \times 
\frac{i}{\mathbf{D}_0} + \dots \label{pertexp}
\end{equation}
The first order perturbation is the second term on the right hand side of (\ref{pertexp}).

Let us define the kinetic operator of a massless, minimally coupled scalar as 
$D_A \equiv \partial_{\mu} a^{D-2} \partial^{\mu}$. The graviton kinetic operator in 
our gauge (\ref{newgauge}) is, 
\begin{eqnarray}
\lefteqn{\mathcal{D}^{\mu\nu\rho\sigma} = \frac12 \Bigl[ \eta^{\mu (\rho} 
\eta^{\sigma )\nu} - \Bigl(\frac{2 \alpha \!-\! \beta^2}{2\alpha}\Bigr) 
\eta^{\mu\nu} \eta^{\rho\sigma}\Bigr] D_A + \Bigl( \frac{D \!-\! 2}{\alpha}\Bigr)
H^2 a^D \delta^{(\mu}_{~~0} \eta^{\nu ) (\rho} \delta^{\sigma )}_{~~0} }
\nonumber \\
& & \hspace{0cm} - \Bigl( \frac{\alpha \!-\! 1}{\alpha} \Bigr) \partial^{(\rho}
a^{D-2} \eta^{\sigma ) (\mu} \partial^{\nu)} + \Bigl( \frac{\alpha \!-\! \beta}{
2 \alpha}\Bigr) \Bigl[ a^{D-2} \partial^{\mu} \partial^{\nu} \eta^{\rho\sigma}
\!+\! \eta^{\mu\nu} \partial^{\rho} \partial^{\sigma} a^{D-2} \Bigr] . \qquad
\label{kineticfull} 
\end{eqnarray}
Expanding the kinetic operator (\ref{kineticfull}) in $\delta \alpha$ and
$\delta \beta$ gives,
\begin{equation}
\mathcal{D}^{\mu\nu\rho\sigma} = \mathcal{D}^{\mu\nu\rho\sigma}_{0} + \delta \alpha
\!\times\! \mathcal{D}^{\mu\nu\rho\sigma}_{\alpha} + \delta \beta \!\times\!
\mathcal{D}^{\mu\nu\rho\sigma}_{\beta} + O(\delta^2) \; , \label{kineticpert}
\end{equation}
where the first three operators are,
\begin{eqnarray}
\mathcal{D}^{\mu\nu\rho\sigma}_0 & = & \frac12 \eta^{\mu (\rho} \eta^{\sigma ) \nu} 
D_A - \frac14 \eta^{\mu\nu} \eta^{\rho\sigma} D_A + (D \!-\! 2) H^2 a^D 
\delta^{(\mu}_{~~0} \eta^{\nu ) (\rho} \delta^{\sigma)}_{~~0} \; , \qquad 
\label{kineticzero} \\
\mathcal{D}^{\mu\nu\rho\sigma}_{\alpha} & = & -\frac14 \eta^{\mu\nu} \eta^{\rho\sigma}
D_A - (D \!-\! 2) H^2 a^D \delta^{(\mu}_{~~0} \eta^{\nu ) (\rho} \delta^{\sigma )}_{~~0}
\nonumber \\
& & \hspace{1cm} - \partial^{(\rho} a^{D-2} \eta^{\sigma )(\mu} \partial^{\nu )} + 
\frac12 \Bigl[ a^{D-2} \partial^{\mu} \partial^{\nu} \eta^{\rho\sigma} + \eta^{\mu\nu} 
\partial^{\rho} \partial^{\sigma} a^{D-2} \Bigr] \; , \qquad \label{kineticalpha} \\
\mathcal{D}^{\mu\nu\rho\sigma}_{\beta} & = & \frac12 \eta^{\mu\nu} \eta^{\rho\sigma} D_A 
- \frac12 \Bigl[ a^{D-2} \partial^{\mu} \partial^{\nu} \eta^{\rho\sigma} + \eta^{\mu\nu} 
\partial^{\rho} \partial^{\sigma} a^{D-2} \Bigr] \; , \qquad \label{kineticbeta}
\end{eqnarray}

From expression (\ref{pertexp}) the $\delta \alpha$ perturbation of the propagator is,
\begin{equation}
i \Bigl[\mbox{}_{\mu\nu} \Delta^{\alpha}_{\rho\sigma}\Bigr](x;x') = \int \!\! d^Dz
\, i \Bigl[\mbox{}_{\mu\nu} \Delta^{0}_{\alpha\beta}\Bigr](x;z) \times i 
\mathcal{D}^{\alpha\beta\gamma\delta}_{\alpha} \times i \Bigl[\mbox{}_{\gamma\delta} 
\Delta^{0}_{\rho\sigma}\Bigr](z;x') \; . \label{alphadef}
\end{equation}
We next substitute expressions (\ref{simpleprop}) for the 0-th order propagator, and 
(\ref{kineticalpha}) for the $\delta \alpha$ perturbation of the kinetic operator.
After some tedious manipulations the result can be expressed in terms of the integrated
propagators (\ref{Idef}-\ref{Kdef}),
\begin{eqnarray}
\lefteqn{ i \Bigl[\mbox{}_{\mu\nu} \Delta^{\alpha}_{\rho\sigma}\Bigr] = 4(D\!-\!2) 
H^2 \delta^0_{~(\mu} \overline{\eta}_{\nu)(\rho} \delta_{\sigma)}^{~0} I_{BB}
-\frac{4H^2}{D\!-\!2}E_{\mu\nu}E_{\rho\sigma}I_{CC} } \nonumber \\
& & \hspace{-0.5cm} + \frac{4H}{D\!-\!3} \Biggl\{-(D\!-\!2) \overline{\eta}_{\mu\nu}
\delta^0_{~(\rho} \overline{D}_{\sigma)} J_{AB} - \overline{\eta}_{\mu\nu}
E_{\rho\sigma} D_0 J_{AC} + E_{\mu\nu} \delta^0_{~(\rho} \overline{D}_{\sigma)}
J_{CB} \nonumber \\
& & \hspace{-0.5cm} + \frac{E_{\mu\nu} E_{\rho\sigma}}{D\!-\!2} D_0 J_{CC} \Biggr\} 
+ \frac{4H}{D\!-\!3} \Biggl\{-(D\!-\!2) \overline{\eta}_{\rho\sigma} \delta^0_{~(\mu}
\overline{\mathcal{D}}_{\nu)} J_{BA} - E_{\mu\nu} \overline{\eta}_{\rho\sigma}
\mathcal{D}_0 J_{CA} \nonumber \\
& & \hspace{-0.5cm} + \delta^0_{~(\mu} \overline{\mathcal{D}}_{\nu)} E_{\rho\sigma}
J_{BC} \!+\! \frac{E_{\mu\nu} E_{\rho\sigma}}{D\!-\!2} \mathcal{D}_0 J_{CC} \Biggr\}
\!+\! \frac{4 D_0 \mathcal{D}_0} {(D\!-\!3)^2} \Biggl\{ \overline{\eta}_{\mu\nu}
\overline{\eta}_{\rho\sigma} K_{AA} \!+\! E_{\mu\nu} E_{\rho\sigma} K_{CC} \nonumber \\
& & \hspace{-0.5cm} - \overline{\eta}_{\mu\nu} E_{\rho\sigma} K_{AC} \!-\! E_{\mu\nu} 
\overline{\eta}_{\rho\sigma} K_{CA}\Biggr\} \!+\! \frac{4}{D\!-\!3} \Biggl\{ 
\delta^0_{(\rho} \overline{D}_{\sigma)} \mathcal{D}_0 \Bigl[-\overline{\eta}_{\mu\nu} 
K_{AB} \!+\! E_{\mu\nu} K_{CB} \Bigr] \nonumber \\
& & \hspace{-0.5cm} - D_0 \delta^0_{(\mu} \overline{\mathcal{D}}_{\nu)} \Bigl[
\overline{\eta}_{\rho\sigma} K_{BA} \!-\! E_{\rho\sigma} K_{BC} \Bigr] \Biggr\} \!-\!
4 \Biggl\{ \overline{D}_{(\rho} \overline{\eta}_{\sigma)(\mu} \overline{\mathcal{D}}_{\nu)} 
K_{AA} \!+\! \delta^0_{~(\mu} \overline{\eta}_{\nu)(\rho} \delta_{\sigma)}^{~0}
D_0 \mathcal{D}_0 \nonumber \\
& & \hspace{-0.5cm} \times K_{BB} \!-\! \delta^0_{~(\rho} \overline{D}_{\sigma)} 
\delta^0_{~(\mu} \overline{\mathcal{D}}_{\nu)} K_{BB} \!+\! D_0\overline{\mathcal{D}}_{(\mu} 
\overline{\eta}_{\nu)(\rho} \delta_{\sigma)}^{~0} K_{AB} \!+\! \overline{D}_{(\rho} 
\overline{\eta}_{\sigma)(\mu} \delta_{\nu)}^{~0} \mathcal{D}_0 K_{BA} \Biggr\} . \qquad
\label{Deltaalpha}
\end{eqnarray}
Recall that $\overline{\eta}^{\mu\nu} = \eta^{\mu\nu} + \delta^{\mu}_{~0} \delta^{\nu}_{~0}$ 
and $E^{\mu\nu} \equiv (D-3) \delta^{\mu}_{~0} \delta^{\nu}_{~0} + \overline{\eta}^{\mu\nu}$.

By analogy with (\ref{alphadef}), the $\delta \beta$ perturbation is,
\begin{equation}
i \Bigl[\mbox{}_{\mu\nu} \Delta^{\beta}_{\rho\sigma}\Bigr](x;x') = \int \!\! d^Dz
\, i \Bigl[\mbox{}_{\mu\nu} \Delta^{0}_{\alpha\beta}\Bigr](x;z) \times i 
\mathcal{D}^{\alpha\beta\gamma\delta}_{\beta} \times i \Bigl[\mbox{}_{\gamma\delta} 
\Delta^{0}_{\rho\sigma}\Bigr](z;x') \; . \label{betadef}
\end{equation}
Substituting the 0-th order propagator (\ref{simpleprop}) and the $\delta \beta$
perturbation of the knietic operator (\ref{kineticbeta}) gives,
\begin{eqnarray}
\lefteqn{i \Bigl[\mbox{}_{\mu\nu} \Delta^{\beta}_{\rho\sigma}\Bigr] = -\frac{4(D\!-\!1)H^2}{
(D \!-\! 3)^2} \Biggl\{ (D \!-\! 2) \overline{\eta}_{\mu\nu} \overline{\eta}_{\rho\sigma}
I_{AA} - \Bigl[ \overline{\eta}_{\mu\nu} E_{\rho\sigma} \!+\! E_{\mu\nu}
\overline{\eta}_{\rho\sigma}\Bigr] I_{AC} } \nonumber \\
& & \hspace{-0.5cm} + \frac{ E_{\mu\nu} E_{\rho\sigma} I_{CC}}{D-2}
\Biggr\} -\frac{4 \overline{\eta}_{\mu\nu}}{D \!-\!3} \Biggl\{ \overline{D}_{\rho}
\overline{D}_{\sigma} K_{AA} + 2 \delta^0_{~(\rho} \overline{D}_{\sigma)} D_0 K_{AB} +
\delta^0_{~\rho} \delta^0_{~\sigma} D_0^2 K_{AC} \nonumber \\
& & \hspace{-0.5cm} + \frac{\overline{\eta}_{\rho\sigma} D_0^2 (K_{AC} \!-\! K_{AA})}{D-3} 
\Biggr\} + \frac{4 E_{\mu\nu}}{(D \!-\!3) (D\!-\!2)} \Biggl\{\overline{D}_{\rho}
\overline{D}_{\sigma} K_{CA} + 2 \delta^0_{~(\rho} \overline{D}_{\sigma)}D_0 K_{CB} 
\nonumber \\
& & \hspace{-0.5cm} + \delta^0_{~\rho} \delta^0_{~\sigma} D_0^2 K_{CC} + 
\frac{ \overline{\eta}_{\rho\sigma} D_0^2 (K_{CC} \!-\! K_{CA})}{D - 3} \Biggr\}
-\frac{4 \overline{\eta}_{\rho\sigma}}{D \!-\!3} \Biggl\{ \overline{\mathcal{D}}_{\mu}
\overline{\mathcal{D}}_{\nu} K_{AA} \nonumber \\
& & \hspace{-0.5cm} + 2 \delta^0_{~(\mu} \overline{\mathcal{D}}_{\nu)} \mathcal{D}_0 
K_{BA} \!+\! \delta^0_{~\mu} \delta^0_{~\nu} {\mathcal{D}_0}^2 K_{CA} \!+\! 
\frac{\overline{\eta}_{\mu\nu} {\mathcal{D}_0}^2 (K_{CA} \!-\! K_{AA})}{D-3} \Biggr\} 
\!+\! \frac{4 E_{\rho\sigma}}{(D \!-\!3) (D\!-\!2)} \nonumber \\
& & \hspace{-0.7cm} \times \!\Biggl\{\overline{\mathcal{D}}_{\mu} \overline{\mathcal{D}}_{\nu} 
K_{AC} \!+\! 2 \delta^0_{~(\mu} \overline{\mathcal{D}}_{\nu)}\mathcal{D}_0 K_{BC} \!+\!
\delta^0_{~\mu} \delta^0_{~\nu} {\mathcal{D}_0}^2 K_{CC} \!+\! \frac{ \overline{\eta}_{\mu\nu}
{\mathcal{D}_0}^2 (K_{CC} \!-\! K_{AC})}{D - 3} \!\Biggr\} . \quad \label{Deltabeta}
\end{eqnarray}
Unlike the $\delta \alpha$ perturbation (\ref{Deltaalpha}), the $\delta \beta$ perturbation
has only diagonal tensor factors and involves no $J_{\mu\nu}(x;x')$ integrals.

\section{Discussion} \label{discuss}

This paper concerns gauge dependence in the graviton propagator on de Sitter 
background. In section~\ref{2pgauge} we generalized the simple gauge condition 
(\ref{simplegauge}) to a 2-parameter family of gauges (\ref{newgauge}). The flat space
limit of this family coincides with the gauges employed in a study of how source and
observer corrections cancel gauge dependence in the effective field equation
\cite{Miao:2017feh}. In section~\ref{solution} we argued that just the first order
perturbations around the simple gauge provide two thirds of the checks made in that
study. Our results for these first order perturbations are equations (\ref{Deltaalpha})
and (\ref{Deltabeta}). They are expressed in terms of integrated propagators that
are evaluated in the Appendix. All our work was done in $D$ spacetime dimensions to
facilitate the use of dimensional regularization.

Even the first order perturbations are very complicated, and we do not advocate using
them for routine calculations. Their purpose is to provide an explicit check that our
technique for canceling gauge dependence \cite{Miao:2017feh} works on de Sitter 
background. Establishing that fact is crucial, but once it has been done, we expect
that future computations will be made using the simple gauge propagator \cite{Tsamis:1992xa,
Woodard:2004ut}. The advantages of this propagator are that its $D=4$ limit is simple, 
that its tensor factors are independent of space and time, and that the 1PI 2-point 
functions it produces are elementary functions of the scale factors and the conformal 
coordinate interval. This is why just one loop computation has been done in another 
gauge, and that computation was only made to check gauge dependence. 

The chief source of complication in our analysis is the factors of $1/a$ and $1/a^2$ 
that accompany derivatives inside convolutions of propagators (\ref{Jdef}-\ref{Kdef}). 
This becomes apparent in the Appendix, when comparing the trivial result 
(\ref{Imunuzero}-\ref{Imunuone}) for the $I_{\mu\nu}(x;x')$ convolution, which contains 
no factors of $1/a$, with the terrific effort expended to determine the $J_{\mu\nu}(x;x')$ 
and $K_{\mu\nu}(x;x')$ convolutions. In addition to facilitating checks of gauge 
dependence, this work should prove useful for computing the expectation values of gauge 
invariant measures of back-reaction \cite{Miao:2017vly}. Were we to push one order 
higher in the perturbations $\delta \alpha$ and $\delta \beta$ there would be 
another convolution, but no more factors of $1/a$ within any one integration, so perhaps 
the extra labor would not be prohibitive. Note that including even one quadratic 
perturbation in flat space would suffice to cover the full range of checks available in 
Table~\ref{Cab} from the all-orders result. Further, only two convolutions are required 
for the full propagator (\ref{flatgrav}) in flat space, so perhaps an all-orders result 
could be obtained as well in de Sitter.

Finally, we should lay out our program of research concerning gauge dependence
on de Sitter background:
\begin{enumerate}
\item{Use the one graviton loop contribution to the 1PI 2-point function of a 
massless, minimally coupled scalar \cite{Kahya:2007bc} to check for logarithmic
corrections to the mode functions and the scalar exchange potential.}
\item{Use the simple propagator to include source and observer corrections to see how 
large logarithms are affected.}
\item{Re-do 1-2 using (\ref{Deltaalpha}) and (\ref{Deltabeta}) to check that source and
observer corrections cancel gauge dependence.}
\end{enumerate}

\vskip 0.5cm

\centerline{\bf Acknowledgements}

This work was partially supported by the Fonds de la Recherche Scientifique
-- FNRS under Grant IISN 4.4517.08 -- Theory of fundamental interactions, 
by Taiwan MOST grants 103-2112-M-006-001-MY3 and 107-2119-M-006-014; by the 
D-ITP consortium, a program of the Netherlands Organization for Scientific 
Research (NWO) that is funded by the Dutch Ministry of Education, Culture 
and Science (OCW); by NSF grants PHY-1806218 and PHY-1912484; and by the 
Institute for Fundamental Theory at the University of Florida.

\section{Appendix: Integrated Propagators} \label{append}

This appendix concerns technical details of the three integrated 
propagators (\ref{Idef}-\ref{Kdef}). We begin by explaining how to 
reflect derivatives from one argument of a propagator to the other. 
We then give exact results for $I_{\mu\nu}(x;x')$, and derive expansions 
for $J_{\mu\nu}(x;x')$ and $K_{\mu\nu}(x;x')$.

\subsection{Reflecting Derivatives}

Although we choose to keep the derivatives of (\ref{DJ}-\ref{DDK}) on the
dummy variable $z^{\mu}$, they could be reflected to the external variables
using some identities which were originally derived in \cite{Tsamis:1992zt}.
Of course space derivatives reflect the same as on flat space background,
\begin{equation}
\partial_i \, i \Delta_{\nu}(x;x') = - \partial'_i \, i \Delta_{\nu}(x;x') \; .
\end{equation}
Reflecting the time derivatives requires that we explain the relation between 
the index $\nu$ in the scalar propagator $i \Delta_{\nu}(x;x')$ and the scalar 
mass $m$,
\begin{equation}
\nu = \sqrt{ \frac{(D \!-\! 1)^2}{4} - \frac{m^2}{H^2} } \; . \label{nutom}
\end{equation}
The three scalar propagators we employ correspond to masses and indices,
\begin{eqnarray}
A & \Longrightarrow & m_A^2 = 0 \qquad , \qquad \nu_A = \Bigl( \frac{D\!-\!1}2
\Bigr) \; , 
\label{nuA}
\\
B & \Longrightarrow & m_B^2 = (D \!-\! 2) H^2 \qquad , \qquad \nu_B = 
\Bigl( \frac{D\!-\!3}2 \Bigr) = \nu_A - 1 \; , 
\label{nuB}
\\
C & \Longrightarrow & m_C^2 = 2 (D \!-\! 3) H^2 \qquad , \qquad \nu_C = 
\Bigl( \frac{D\!-\!5}2 \Bigr) = \nu_B - 1 \; .
\label{nuC}
\end{eqnarray}
The temporal reflection identities are,
\begin{eqnarray}
\Bigl[ \partial_0 + (\nu_A \!-\! \nu) H a\Bigr] i \Delta_{\nu}(x;x') & = &
-\Bigl[ \partial'_0 + (\nu_A \!+\! \nu \!-\! 1) H a'\Bigr] 
i \Delta_{\nu-1}(x;x') \; , \qquad \\
\Bigl[ \partial_0 + (\nu_A \!+\! \nu) H a\Bigr] i \Delta_{\nu}(x;x') & = &
-\Bigl[ \partial'_0 + (\nu_A \!-\! \nu \!-\! 1) H a'\Bigr] 
i \Delta_{\nu+1}(x;x') \; . \qquad 
\end{eqnarray}
The specific reflection identities we need are,
\begin{eqnarray}
\partial_0 i \Delta_{A}(x;x') & = & -\Bigl[ \partial'_0 + (D \!-\! 2) H a'\Bigr] 
i \Delta_{B}(x;x') \; , \qquad 
\label{reflection1}
\\
(\partial_0 \!+\! H a) i \Delta_{B}(x;x') & = & -\Bigl[ \partial'_0 + (D \!-\! 3) 
H a'\Bigr] i \Delta_{C}(x;x') \; . \qquad 
\label{reflection2}
\end{eqnarray}

\subsection{Results}

Exact results can be derived for $I_{\mu\nu}(x;x')$ \cite{Miao:2011fc},
\begin{eqnarray}
I_{\mu\nu}(x;x') &=& \frac{ i\Delta_{\mu}(x;x') \!-\! i\Delta_{\nu}(x;x')}{m_{\mu}^2
\!-\! m_{\nu}^2} \; , \label{Imunuzero}
\qquad \quad
\mu\neq \nu \, ,
\\
I_{\mu\mu}(x;x') &=& \frac{-1}{2\mu H^2} \frac{\partial}{\partial \mu} 
	i \Delta_{\mu}(x;x') \; , \label{Imunuone}
\end{eqnarray}
where from now on we use a shorthand notation~$\nu_\mu\!\equiv\!\mu$.
Although an asymptotic expansion can be derived for $J_{\mu\nu}(x;x')$ 
\cite{Miao:2017vly}, it is better, for our purposes, to use the reflection
identities~(\ref{reflection1}) and~(\ref{reflection2}) to express all the 
$J$-propagators in terms of~derivatives of the~$K$-propagators,
\begin{eqnarray}
J_{AA} &=& 
	\frac{1}{(D\!-\!2)H} \biggl\{
	\Bigl[ \partial_0 + (D\!-\!2) Ha \Bigr] K_{BA}
		+ \Bigl[ \partial_0' + (D\!-\!2) H a' \Bigr] K_{AB}
	\biggr\} \, ,
\label{JAA}
\\
J_{BB} &=&
	\frac{-1}{(D\!-\!2)H} \biggl\{
	\partial_0 K_{AB}
	+ \partial_0' K_{BA}
	\biggr\} \, ,
\label{JBB}
\\
J_{CC} &=&
	\frac{-1}{(D\!-\!4) H} \biggl\{
	\Bigl[ \partial_0 + H a \Bigr] K_{BC}
	+ \Bigl[ \partial_0' + H a' \Bigr] K_{CB}
	\biggr\} \, ,
\label{JCC}
\\
J_{AB} &=&
	\frac{1}{(D\!-\!3)H} \biggl\{
	\Bigl[ \partial_0 + (D\!-\!2) Ha  \Bigr] K_{BB}
	+ \Bigl[ \partial_0' + (D\!-\!3) H a' \Bigr] K_{AC}
	\biggr\} \, , \qquad
\label{JAB}
\\
J_{AC} &=&
	\frac{1}{H} \biggl\{
	\Bigl[ \partial_0 + (D\!-\!2) Ha \Bigr] K_{BC}
	+ \Bigl[ \partial_0' + Ha' \Bigr] K_{AB}
	\biggr\} \, ,
\label{JAC}
\\
J_{BC} &=&
	\frac{-1}{(D\!-\!3)H} \biggl\{
		\partial_0 K_{AC} 
		+ \Bigl[ \partial_0' \!+\! Ha' \Bigr] K_{BB}
	\biggr\} \, .
\label{JBC}
\end{eqnarray}
Thus we only need to compute the simpler~$K$-propagators, which we do in the remainder
of the appendix. 
If necessary, analogous relations expressing~$I$-propagators in terms of the
derivatives of the~$J$-propagators can be constructed and used to check the correctness
of the solutions.

We determine the~$K$-propagators by solving the two equations of motion they satisfy,
\begin{eqnarray}
\Bigl( \square - M_\mu^2 \Bigr) K_{\mu\nu}(x;x')
	&=& \frac{1}{a^2} \times i \Delta_\nu(x;x') \, ,
\label{K eqs1}
\\
\Bigl( \square' - M_\nu^2 \Bigr) K_{\mu\nu}(x;x')
	&=& \frac{1}{a'^2} \times i \Delta_\mu(x;x') \, .
\label{K eqs2}
\end{eqnarray}
Instead of the two coordinates~$x$ and~$x'$, it is convenient to use different variables --
the de Sitter invariant distance~$y$ defined in~(\ref{ydef}), and the two-time 
variables~$u\!=\! \ln(aa')$ and~$v \!=\! \ln(a/a')$. Furthermore, it is convenient to rescale
the~$K$-propagators,
\begin{equation}
K_{\mu\nu}(x;x') = \frac{H^{D-2} \, \Gamma\bigl( \frac{D-2}{2} \bigr) }{(4\pi)^{D/2}}
	\times \frac{e^{-u}}{H^2} \times \mathcal{K}_{\mu\nu}(y,u,v) \, ,
\end{equation}
and to consider the difference and the sum of the equations~(\ref{K eqs1}) and~(\ref{K eqs2}),
respectively,
\begin{eqnarray}
&&
\Biggl\{
	8 \frac{\partial}{\partial v} \biggl[ {\rm sh}^2 \Bigl( \frac{v}{2} \Bigr) \frac{\partial}{\partial y} \biggr]
	- 4 \, {\rm sh}(v) \frac{\partial}{\partial u} \frac{\partial}{\partial y}
\nonumber \\
&&	\hspace{1cm}
	+ \biggl[ 2 y \frac{\partial}{\partial y} \!+\! 2 \frac{\partial}{\partial u} \!+\! D\!-\!3 \biggr]
		\frac{\partial}{\partial v}
	- \frac{(\mu^2 \!-\! \nu^2)}{2}
	\Biggr\} \mathcal{K}_{\mu\nu}(y,u,v)
\nonumber \\
&&	\hspace{4.5cm}
	=  {\rm sh}(v) \, i \Delta_{\mu\nu}^+(y,u)
	+ {\rm ch}(v) \, i \Delta_{\mu\nu}^-(y,u) \, ,
\label{diff eq}
\\
&&
\Biggl\{ \biggl[ (4 y \!-\! y^2) \frac{\partial}{\partial y} 
	- \biggl( D \!-\! 2 \!+\! 2\frac{\partial}{\partial u} \biggr) y
	- 8 \, {\rm sh}^2\bigl( \frac{v}{2} \bigr) \frac{\partial}{\partial u}
		+ 2 (D\!-\!2) \biggr] \frac{\partial}{\partial y}
\nonumber \\
&&	\hspace{1cm}	- \frac{\partial^2}{\partial v^2} 
	+ 4 \frac{\partial}{\partial v} \biggl[ {\rm sh}(v) \frac{\partial}{\partial y} \biggr]
	- \frac{\partial^2}{\partial u^2}
	- (D\!-\!3) \frac{\partial}{\partial u}
\nonumber \\
&&	\hspace{2cm}
		+ \frac{(\mu^2 \!+\! \nu^2)}{2} - \frac{(D\!-\!3)^2}{4} \Biggr\} \mathcal{K}_{\mu\nu}(y,u,v)
\nonumber \\
&& \hspace{4.5cm}
	= {\rm ch}(v) \, i\Delta_{\mu\nu}^+(y,u)
	+ {\rm sh}(v) \, i\Delta_{\mu\nu}^-(y,u) \, , \qquad \qquad
\label{sum eq}
\end{eqnarray}
where the sources on the right hand side are rescaled sums and differences of the
scalar propagators~(\ref{Adef}-\ref{Cdef}), defined as,
\begin{equation}
i \Delta_{\mu\nu}^\pm(y,u) =
	\Biggl[ \frac{H^{D-2} \, \Gamma\bigl( \frac{D-2}{2} \bigr) }{(4\pi)^{D/2}} \Biggr]^{-1}
	\times \frac{1}{2}
		\biggl[ i \Delta_\mu(x;x') \pm i \Delta_\nu(x;x') \biggr] \, .
\label{S+-def}
\end{equation}
We solve the equations~(\ref{diff eq}) and~(\ref{sum eq}) 
as a power series in~$y$ around~$y\!=\!0$.
In order to satisfy the original equations~(\ref{K eqs1})
and~(\ref{K eqs2}), this power series has to take the form,
\begin{eqnarray}
\mathcal{K}_{\mu\nu}(y,u,v)
		&=& - \frac{2(E_{\mu\nu})_0 }{(D\!-\!4)} \Bigl( \frac{y}{4} \Bigr)^{- \frac{D-4}{2}}
		+ \frac{2 (F_{\mu\nu})_0}{(D\!-\!4)}
\nonumber \\
&&
	+ \sum_{n=1}^{\infty} \biggl[
		\frac{(E_{\mu\nu})_n}{n - \frac{D-4}{2}} \Bigl( \frac{y}{4} \Bigr)^{n - \frac{D-4}{2}}
		-
		\frac{(F_{\mu\nu})_n}{n} \Bigl( \frac{y}{4} \Bigr)^{n}
		\biggr]  \, ,
\label{K ansatz}
\end{eqnarray}
where numerical factors have been taken out for convenience, and
where the coefficients~$(E_{\mu\nu})_n$ and~$(F_{\mu\nu})_n$ that we need to
determine are functions of variables~$u$ and~$v$ in general. The equations of motion for
these coefficients follow from the full equations~(\ref{diff eq}) and~(\ref{sum eq}).
It is useful to define an analogous power series expansion of the
sum and the difference of the scalar propagators from~(\ref{S+-def}),
\begin{eqnarray}
i\Delta_{\mu\nu}^+(x;x') &=& 
	\Bigl( \frac{y}{4} \Bigr)^{- \frac{D-2}{2}}
	+ \sum_{n=0}^{\infty} \biggl[ (S_{\mu\nu}^+)_n \Bigl( \frac{y}{4} \Bigr)^{n - \frac{D-4}{2}} 
		- (Q_{\mu\nu}^+)_n \Bigl( \frac{y}{4} \Bigr)^{n } \biggr] \, , \quad
\\
i\Delta_{\mu\nu}^-(x;x') &=& 
	\sum_{n=0}^{\infty} \biggl[ (S_{\mu\nu}^-)_n \Bigl( \frac{y}{4} \Bigr)^{n - \frac{D-4}{2}} 
		- (Q_{\mu\nu}^-)_n \Bigl( \frac{y}{4} \Bigr)^{n } \biggr] \, ,
\end{eqnarray}
where the coefficients of these expansions can be read off by comparing~(\ref{S+-def}) 
to the power series of the scalar
propagators~(\ref{Adef}-\ref{Cdef}),
\begin{eqnarray}
(S_{\mu\nu}^\pm)_n &=&
	\frac{\Gamma\bigl( \frac{4-D}{2} \bigr)}{ 2(n\!+\!1)! \, \Gamma\bigl( \frac{6-D}{2} \!+\! n \bigr)}
	\hspace{5cm} (n\ge0)
\nonumber \\
&&	\hspace{0.cm}
	\times 
	\biggl[ \frac{\Gamma\bigl( \frac{3}{2} \!+\! \mu \!+\! n \bigr) \,
		\Gamma\bigl( \frac{3}{2} \!-\! \mu \!+\! n \bigr) }
			{\Gamma\bigl( \frac{1}{2} \!+\! \mu \bigr) \,
				\Gamma\bigl( \frac{1}{2} \!-\! \mu \bigr)}
		\pm
		\frac{\Gamma\bigl( \frac{3}{2} \!+\! \nu \!+\! n \bigr) \,
		\Gamma\bigl( \frac{3}{2} \!-\! \nu \!+\! n \bigr) }
			{\Gamma\bigl( \frac{1}{2} \!+\! \nu \bigr) \,
				\Gamma\bigl( \frac{1}{2} \!-\! \nu \bigr)}
		\biggr] \, ,
\\
(Q_{AA}^\pm)_0 &=&
	\frac{\Gamma(D\!-\!1)}{2 \, \Gamma\bigl( \frac{D}{2} \bigr) \, \Gamma\bigl( \frac{D-2}{2} \bigr)}
	\biggl[ - \frac{A_1 \, \Gamma\bigl( \frac{D}{2} \bigr)}{\Gamma(D\!-\!1)} - u \biggr]
	\bigl( 1 \pm 1 \bigr) \, ,
\label{QAA}
\\
(Q_{BB}^\pm)_0 &=&
	\frac{\Gamma(D\!-\!2)}{2 \, \Gamma\bigl( \frac{D}{2} \bigr) \, 
		\Gamma\bigl( \frac{D-2}{2} \bigr) } \bigl( 1 \pm 1 \bigr) \, ,
\label{QBB}
\\
(Q_{CC}^\pm)_0 &=&
	- \frac{\Gamma(D\!-\!3)}{2 \, \Gamma\bigl( \frac{D}{2} \bigr) \, 
		\Gamma\bigl( \frac{D-2}{2} \bigr) } \bigl( 1 \pm 1 \bigr) \, ,
\label{QCC}
\\
(Q_{AB}^\pm)_0 &=&
	\frac{\Gamma(D\!-\!1)}{2 \, \Gamma\bigl( \frac{D}{2} \bigr) \, \Gamma\bigl( \frac{D-2}{2} \bigr)}
	\biggl[ - \frac{A_1 \, \Gamma\bigl( \frac{D}{2} \bigr)}{\Gamma(D\!-\!1)}
	\pm \frac{1}{(D\!-\!2)} - u \biggr] \, ,
\label{QAB}
\\
(Q_{AC}^\pm)_0 &=&
	\frac{\Gamma(D\!-\!1)}{2 \, \Gamma\bigl( \frac{D}{2} \bigr) \, \Gamma\bigl( \frac{D-2}{2} \bigr)}
	\biggl[ - \frac{A_1 \, \Gamma\bigl( \frac{D}{2} \bigr)}{\Gamma(D\!-\!1)}
	\mp \frac{1}{(D\!-\!2)(D\!-\!3)} - u \biggr] \, ,
\label{QAC}
\\
(Q_{BC}^\pm)_0 &=&
	\frac{\Gamma( D\!-\!3) }{ 2 \, \Gamma\bigl( \frac{D}{2} \bigr) \, \Gamma\bigl( \frac{D-2}{2} \bigr)}
	\bigl( D-3 \mp 1 \bigr)
\label{QBC}
\\
(Q_{\mu\nu}^\pm)_n &=&
	\frac{\Gamma\bigl( \frac{4-D}{2} \bigr)}{2 \, n!\,\Gamma\bigl( \frac{D}{2} \!+\! n \bigr)}
\hspace{6cm} (n\ge1)
\nonumber \\
&&	\hspace{-1.5cm}
	\times \biggl[
	\frac{\Gamma\bigl( \frac{D-1}{2} \!+\! n\!+\! \mu \bigr) \,
			\Gamma\bigl( \frac{D-1}{2} \!+\! n \!-\! \mu \bigr) }
				{\Gamma\bigl( \frac{1}{2} \!+\! \mu \bigr) \,
					\Gamma\bigl( \frac{1}{2} \!-\! \mu \bigr)}
	\pm
	\frac{\Gamma\bigl( \frac{D-1}{2} \!+\! n\!+\! \nu \bigr) \,
			\Gamma\bigl( \frac{D-1}{2} \!+\! n \!-\! \nu \bigr) }
				{\Gamma\bigl( \frac{1}{2} \!+\! \nu \bigr) \,
					\Gamma\bigl( \frac{1}{2} \!-\! \nu \bigr)}
	\biggr] \, , \qquad
\label{Qgen}
\end{eqnarray}
where the constant~$A_1$ is,
\begin{equation}
A_1 = -\frac{\Gamma(D\!-\!1)}{\Gamma\bigl( \frac{D}{2} \bigr)}
	\pi {\rm cot}\Bigl(\pi \frac{D}{2}\Bigr) \, ,
\end{equation}
and where~$\psi(z)\!=\!\Gamma'(z)/\Gamma(z)$ is the digamma function.
Note that some coefficients from the integer
series depend linearly on~$u\!=\!\ln(aa')$, and that this dependence 
descends from the de Sitter breaking part of the $A$-type propagator~(\ref{Adef}).

Since the equations for the coefficients of the non-integer power 
series~(\ref{K ansatz}) decouple from the equations for the integer ones, 
we solve for them separately in the following subsections.

\subsubsection{Non-integer power series}

The equations for the non-integer power series coefficients, descending 
from Eq.~(\ref{diff eq}) are,
\begin{eqnarray}
&&
2 \frac{\partial}{\partial v} \Biggl[ {\rm sh}^2\Bigl( \frac{v}{2} \Bigr) (E_{\mu\nu})_0 \Biggr]
	- {\rm sh}(v) \frac{\partial}{\partial u} (E_{\mu\nu})_0 = {\rm sh}(v) \, ,
\label{D diff 0}
\\
&&
2 \frac{\partial}{\partial v} \Biggl[ {\rm sh}^2\Bigl( \frac{v}{2} \Bigr) (E_{\mu\nu})_n \Biggr]
	- {\rm sh}(v) \frac{\partial}{\partial u} (E_{\mu\nu})_n
\hspace{4cm} (n \ge 1) 
\nonumber \\
&&	\hspace{2cm}
	= \frac{-1}{ \bigl(n \!-\! \frac{D-2}{2} \bigr)}	\Biggl[
	\biggl( 2n \!-\! 1 \!+\! 2 \frac{\partial}{\partial u} \biggr) \frac{\partial}{\partial v} 
		- \frac{(\mu^2 \!-\! \nu^2)}{2} 
	\Biggr] (E_{\mu\nu})_{n-1} 
\nonumber \\
&&	\hspace{3cm}
	+ \,{\rm sh}(v) \, (S_{\mu\nu}^+)_{n-1} + {\rm ch}(v) \, (S_{\mu\nu}^-)_{n-1} \, . 
\label{D diff n}
\end{eqnarray}
while the ones descending from Eq.~(\ref{sum eq}) are,
\begin{eqnarray}
&&
\frac{\partial}{\partial v} \biggl[ {\rm sh}(v) (E_{\mu\nu})_0 \biggr]
	- 2 \, {\rm sh}^2\Bigl( \frac{v}{2} \Bigr) \frac{\partial}{\partial u} (E_{\mu\nu})_0
	= {\rm ch}(v) \, ,
\label{D sum 0}
\\
&&
\frac{\partial}{\partial v} \biggl[ {\rm sh}(v) (E_{\mu\nu})_n \biggr]
	- 2 \, {\rm sh}^2\Bigl(\frac{v}{2} \Bigr) \frac{\partial}{\partial u} (E_{\mu\nu})_n + n (E_{\mu\nu})_n
\hspace{2cm} (n\ge1)
\nonumber \\
&&	\hspace{0.5cm}	
	= \frac{1}{\bigl(n \!-\! \frac{D-2}{2} \bigr) } \Biggl[
		\biggl( n \!-\! \frac{D\!-\!2}{2} \biggr) 
		\biggl( n \!+\! \frac{D\!-\!4}{2} \!+\! 2 \frac{\partial}{\partial u} \biggr)
		+ \frac{\partial^2}{\partial v^2}
\nonumber \\
&&	\hspace{3.1cm}
		+ \frac{\partial^2}{\partial u^2}
		+ (D\!-\!3) \frac{\partial}{\partial u}
		+ \frac{(D\!-\!3)^2}{4} 
		- \frac{(\mu^2 \!+\! \nu^2)}{2}
		\Biggr] (E_{\mu\nu})_{n-1}
\nonumber \\
&&	\hspace{5.5cm}
		+ \, {\rm ch}(v) \, (S_{\mu\nu}^+)_{n-1}
		+ {\rm sh}(v) \, (S_{\mu\nu}^-)_{n-1} \, . \qquad 
\label{D sum n}
\end{eqnarray}

The leading coefficient that simultaneously solves~(\ref{D diff 0})
and~(\ref{D sum 0}), and is finite in the time coincidence limit ({\it i.e.} limit~$v\!\to\!0$) is
unique,
\begin{equation}
(E_{\mu\nu})_0 = 1 \, ,
\label{E_0}
\end{equation}
and is independent of the indices~($\mu,\nu=A,B,C$). This solution is
 taken as the germ of the recurrence in~(\ref{D diff n}), which can be easily 
integrated as it is only a first order differential equation. Iterating this equation generates
higher coefficients of the expansion, and all the integration constants are 
uniquely fixed by demanding expressions to be finite for~$v\!\to\!0$. The first couple of 
higher order coefficients are,
\begin{eqnarray}
(E_{\mu\nu})_1 &=& 
	(S_{\mu\nu}^+)_{0}
	+ \frac{(\mu^2 \!-\! \nu^2 )}{2(D\!-\!4)}
		\Biggl[ \frac{{\rm sh}(v) \!-\! v}{{\rm sh}^2\bigl( \frac{v}{2} \bigr) } \Biggr]
\label{E_1}
\, ,
\\
(E_{\mu\nu})_2 &=& 
	(S_{\mu\nu}^+)_1
	+ \frac{3(\mu^2 \!-\! \nu^2)}{2(D\!-\!4)(D\!-\!6)}
		\Biggl[ \frac{3{\rm sh}(v) - v \bigl[ 1 \!+\! 2 \,{\rm ch}(v) \bigr]}{3{\rm sh}^4\bigl( \frac{v}{2} \bigr)} \Biggr]
\nonumber
\\
&& \hspace{1cm}
	+\frac{ (\mu^2 \!-\! \nu^2)}{2 (D\!-\!4)(D\!-\!6)} 
		\Biggl[ \frac{2 \, {\rm sh}\bigl( \frac{v}{2} \bigr) - v \, {\rm ch}\bigl( \frac{v}{2} \bigr) }
				{{\rm sh}^3 \bigl( \frac{v}{2} \bigr) } \Biggr]
\nonumber \\
&&	\hspace{2cm}
	+ \frac{(\mu^2 \!-\! \nu^2) (\mu^2 \!+\! \nu^2 \!-\! \frac{5}{2})}{2 (D\!-\!4)(D\!-\!6)}
		\Biggl[ \frac{{\rm sh}(v) \!-\! v}{{\rm sh}^2\bigl( \frac{v}{2} \bigr) } \Biggr] \, .
\label{E_2}
\end{eqnarray}
Specialization of the expressions above to particular types~$A,B,C$ is
simply accomplished by plugging in the specific parameters~(\ref{nuA}-\ref{nuC}).
As opposed to the general case where the coefficients take complicated forms above,
in the special case of diagonal propagators~($\nu\!=\! \mu$) we can solve the 
recurrence in a closed form to all orders,
\begin{equation}
(E_{\mu\mu})_0 = 1 \, ,
\qquad \qquad \quad
(E_{\mu\mu})_n = (S_{\mu\mu}^+)_{n-1} \, ,
\qquad (n\ge1) \, .
\end{equation}

\subsubsection{Integer power series}

The equations for the coefficients of the integer power series in~(\ref{K ansatz})
that descend from~(\ref{diff eq}) are,
\begin{eqnarray}
&&
2 \frac{\partial}{\partial v} \Biggl[ {\rm sh}^2 \Bigl( \frac{v}{2} \Bigr) (F_{\mu\nu})_1 \Biggr]
	- {\rm sh}(v) \frac{\partial}{\partial u} (F_{\mu\nu})_1
\nonumber \\
&&	\hspace{1.5cm}
	- \frac{2}{(D\!-\!4)} \Biggl[ \biggl( 2 \frac{\partial}{\partial u} \!+\! D\!-\!3 \biggr) 
		\frac{\partial}{\partial v} - \frac{(\mu^2 \!-\! \nu^2)}{2} \Biggr] (F_{\mu\nu})_0
\nonumber \\
&&	\hspace{5cm}
	= {\rm sh}(v) \, (Q_{\mu\nu}^+)_0 + {\rm ch}(v) \, (Q_{\mu\nu}^-)_0 \, ,
\label{F diff 0}
\\
&&
2 \frac{\partial}{\partial v} \Biggl[ {\rm sh}^2 \Bigl( \frac{v}{2} \Bigr) (F_{\mu\nu})_n \Biggr]
	- {\rm sh}(v) \frac{\partial}{\partial u} (F_{\mu\nu})_n
\hspace{4cm} (n\ge2)
\nonumber \\
&&	\hspace{1.5cm}
	= \frac{-1}{(n\!-\!1)} \Biggl[ 
		\biggl( 2n\!+\! D\!-\!5 \!+\! 2 \frac{\partial}{\partial u} \biggr) \frac{\partial}{\partial v}
			- \frac{(\mu^2 \!-\! \nu^2)}{2} \Biggr] (F_{\mu\nu})_{n-1}
\nonumber \\
&&	\hspace{4cm}
	+ {\rm sh}(v) \, (Q_{\mu\nu}^+)_{n-1}
	+ {\rm ch}(v) \, (Q_{\mu\nu}^-)_{n-1} \, ,
\label{F diff n}
\end{eqnarray}
and the equations that descend from~(\ref{sum eq}),
\begin{eqnarray}
&&\hspace{-0.8cm}
\frac{\partial}{\partial v} \biggl[ {\rm sh}(v) (F_{\mu\nu})_1 \biggr]
	+ \Biggl[ \frac{D\!-\!2}{2} - 2 \, {\rm sh}^2 \bigl( \frac{v}{2} \bigr) \frac{\partial}{\partial u} \Biggr]
		 (F_{\mu\nu})_1
\nonumber \\
&&	\hspace{-0.5cm}
	+ \frac{2}{(D\!-\!4)} \Biggl[
		\frac{\partial^2}{\partial v^2} + \frac{\partial^2}{\partial u^2}
		+ (D\!-\!3) \frac{\partial}{\partial u} + \frac{(D\!-\!3)^2}{4}
		- \frac{(\mu^2\!+\!\nu^2)}{2}
	\Biggr] (F_{\mu\nu})_0
\nonumber \\
&&	\hspace{5cm}
	= {\rm ch}(v) \, (Q_{\mu\nu}^+)_0 + {\rm sh}(v) \, (Q_{\mu\nu}^-)_0 \, , \qquad
\label{F sum 0}
\\
&&\hspace{-0.8cm}
\frac{\partial}{\partial v} \biggl[ {\rm sh}(v) (F_{\mu\nu})_n \biggr]
	+ \Biggl[ n \!+\! \frac{D\!-\!4}{2} \!-\! 2\, {\rm sh}^2 \Bigl( \frac{v}{2} \Bigr) \frac{\partial}{\partial u} \Biggr]
		(F_{\mu\nu})_n
	\hspace{2.5cm} (n\ge2)
\nonumber \\
&&\hspace{-0.cm} =
	\frac{1}{(n\!-\!1)} \Biggl[ \frac{\partial^2}{\partial v^2} + \frac{\partial^2}{\partial u^2}
		+ (2n\!+\! D \!-\! 5) \frac{\partial}{\partial u}
\nonumber \\
&&	\hspace{1cm}
		+ (n\!-\!1) (n\!+\!D\!-\!4)
		+ \frac{(D\!-\!3)^2}{4} - \frac{(\mu^2 \!+\! \nu^2)}{2} \Biggr] (F_{\mu\nu})_{n-1}
\nonumber \\
&&\hspace{2.5cm}
+ {\rm ch}(v) \, (Q_{\mu\nu}^+)_{n-1} + {\rm sh}(v) \, (Q_{\mu\nu}^-)_{n-1} \, .
\label{F sum n}
\end{eqnarray}
In the equations above the first two leading coefficients satisfy two coupled 
equations~(\ref{F diff 0}) and~(\ref{F sum 0}), as opposed to the equations
for non-integer coefficients where only the leading coefficient appears in the leading
order equations. This difference complicates the problem since now
the first two leading coefficients are required to set off the recurrence defining the
higher order coefficients.

For the diagonal propagators we can find the coefficients at all orders in a closed form,
\begin{eqnarray}
&&
(F_{AA})_0 =
	\frac{(D\!-\!4)}{4} (Q_{AA}^+)_0 - \frac{(2D\!-\!3) \, \Gamma(D\!-\!2)}
			{2 \, \Gamma\bigl( \frac{D}{2} \bigr) \, \Gamma\bigl( \frac{D-4}{2} \bigr)}
\, ,
\\
&&	\hspace{2.cm}
	(F_{AA})_1 =
	(Q_{AA}^+)_0 - \frac{\Gamma(D\!-\!1)}
		{\Gamma\bigl( \frac{D}{2} \bigr) \, \Gamma\bigl( \frac{D-2}{2} \bigr)}
	\, ,
\\
&&	\hspace{4.cm}
	(F_{AA})_n =
	(Q_{AA}^+)_{n-1} \, ,
	\qquad \quad
	(n\ge2) \qquad\qquad
\\
&&
(F_{BB})_0 =
	1 - \frac{(D\!-\!2)(D\!-\!4)}{4(D\!-\!3)} (Q_{BB}^+)_0 \, u \, ,
\\
&&	\hspace{2.cm}
(F_{BB})_n =
	(Q_{BB}^+)_{n-1} \, ,
	\qquad \quad
	(n\ge1)
\\
&&
	(F_{CC})_0 =
	- \frac{(D\!-\!2)}{4} (Q_{CC}^+)_0
	+ \frac{1}{2} e^{-(D-4)u}
	\, ,
\\
&&	\hspace{2.cm}
	(F_{CC})_n =
	(Q_{CC}^+)_{n-1} \, .
	\qquad \quad
	(n\ge1)
\end{eqnarray}
The homogeneous parts of the solutions for the leading coefficients were fixed
by requiring that the limit~$D\!\to\!4$ of the full propagator exists off-coincidence.

The off-diagonal coefficients are considerably more difficult to solve for, 
and we give the solutions for the 
coefficients in terms of the power series in~$v$,
\begin{equation}
(F_{\mu\nu})_n = \sum_{k=0}^{\infty} (F_{\mu\nu})_n^k \, v^k \, .
\label{v series}
\end{equation}
This expansion, together with power series in~$y$ from~(\ref{K ansatz}), makes the
answer for the off-diagonal~$K$-propagators take the form of a double power
series in terms less and less singular in the space-time coincidence limit.
The solutions for the coefficients in~(\ref{v series}) are found by expanding the hyperbolic
functions in Eqs.~(\ref{F diff 0}-\ref{F diff n}), and organizing the equations by orders in~$v$.
The two leading equations~(\ref{F diff 0}) and~(\ref{F sum 0}) have to be solved 
first, as they couple only~$(F_{\mu\nu})_0$ and~$(F_{\mu\nu})_1$. We give the
first several terms in the~$v$-expansion of these two in the following form,
\begin{eqnarray}
(F_{\mu\nu})_0^0 &=&
	\frac{(D\!-\!4)}{(\mu^2\!-\!\nu^2)} (Q_{\mu\nu}^-)_0 \qquad , \qquad (F_{\mu\nu})_0^1 = 0
	\, ,
\label{leader}
\\
(F_{\mu\nu})_0^2 &=&
	\frac{- D(D\!-\!4)}{4(D\!-\!1)(D\!-\!2)} 
	\Biggl[ 1 \!-\! \frac{2}{(D\!-\!2)} \frac{\partial}{\partial u} \Biggr]
	\Biggl\{ \frac{1}{D} \biggl[ D \!-\! 2 \!+\! 2 \frac{\partial}{\partial u} \biggr] (Q_{\mu\nu}^+)_0
\nonumber \\
&&	\hspace{-0.5cm}
	+ \biggl( 1 + \frac{\partial}{\partial u} \biggr)
		\Biggl[ (D\!-\!3) \frac{\partial}{\partial u}
			+ \frac{(D\!-\!3)^2}{4} - \frac{(\mu^2 \!+\! \nu^2)}{2} \Biggr]
				\frac{4 (Q_{\mu\nu}^-)_0}{D(\mu^2\!-\!\nu^2)} \Biggr\}
	\, ,	\qquad
\\
(F_{\mu\nu})_1^0 &=&
	\frac{2}{D} \Biggl\{
	- \frac{4(F_{\mu\nu})_0^2}{(D\!-\!4)}  + (Q_{\mu\nu}^+)_0
\nonumber \\
&&	\hspace{1cm}
		- \Biggl[ (D\!-\!3) \frac{\partial}{\partial u}
			+ \frac{(D\!-\!3)^2}{4} - \frac{(\mu^2 \!+\! \nu^2)}{2} \Biggr] 
				\frac{2(Q_{\mu\nu}^-)_0}{(\mu^2\!-\!\nu^2)}
	\Biggr\}  \, ,
\\
(F_{\mu\nu})_0^3 &=&
	\frac{-(D\!+\!2)(D\!-\!4)}{6D(D\!-\!1)} 
		\Biggl[ 1 \!-\! \frac{2}{(D\!-\!1)} \frac{\partial}{\partial u} \Biggr]
\nonumber \\
&&\hspace{0.5cm}
		\times \Biggl\{
		\frac{1}{2} (Q_{\mu\nu}^-)_0
			- \frac{(\mu^2\!-\!\nu^2)}{(D\!-\!4)} (F_{\mu\nu})_0^2
			- \biggl( \frac{3}{2} \!-\! \frac{\partial}{\partial u} \biggr) \frac{2(Q_{\mu\nu}^-)_0}{(D\!+\!2)}
		\Biggr\} \, ,
\\
(F_{\mu\nu})_1^1 &=&
	\frac{2}{(D\!+\!2)} \Biggl[
		(Q_{\mu\nu}^-)_0 -\frac{12}{(D\!-\!4)} (F_{\mu\nu})_0^3
	\Biggr] \, ,
\label{last}
\end{eqnarray}
which is also the order in which they are conveniently solved for.
In the expressions above all the coefficients are assumed to be
(at most) linear functions of~$u$, as is also true for the coefficients~(\ref{QAB}-\ref{Qgen}).
The leading coefficient in~(\ref{leader}) and other homogeneous solutions are chosen such that 
(i) the regular~$(D\!-\!4)$ limit of the full~$K$-propagators exists, (ii)
the subleading coefficients take simpler form. This choice is consistent due to 
the fact that the~$(Q_{\mu\nu}^{\pm})$ coefficients of the three types~$(\mu,\nu) \!=\! (A,B,C)$
satisfy a consistency relation,
\begin{equation}
\biggl[ (D\!-\!1) \frac{\partial}{\partial u}
	+ \frac{(D\!-\!1)^2}{4} - \frac{(\mu^2 \!+\! \nu^2)}{2} \biggr] (Q_{\mu\nu}^-)_0
	+ \frac{(\mu^2\!-\! \nu^2)}{2} (Q_{\mu\nu}^+)_0
	- \frac{D}{2} (Q_{\mu\nu}^-)_1 = 0 \, ,
\end{equation}
which guarantees that the solutions of the leading Eqs.~(\ref{F diff 0})
and~(\ref{F sum 0}) that we found are consistent with higher order equations 
Eqs.~(\ref{F diff n}) and~(\ref{F sum n}).

The higher order coefficients are now easily generated from the 
lower order ones~(\ref{leader}-\ref{last}), by using either of 
equations~(\ref{F diff n}) or~(\ref{F sum n}),~{\it e.g.}, 
\begin{eqnarray}
(F_{\mu\nu})_2^0 &=& 
	\frac{2}{(D\!+\!2)} \Biggl\{
	2 (F_{\mu\nu})_1^2 
	+ (Q_{\mu\nu}^+)_1
\nonumber \\
&&	\hspace{2cm}
	+ \Biggl[ (D\!-\!1) \frac{\partial}{\partial u} + \frac{(D\!-\!1)^2}{4}
		- \frac{(\mu^2 \!+\! \nu^2)}{2} \Biggr] (F_{\mu\nu})_1^0
	\Biggr\} \, ,
\\
(F_{\mu\nu})_2^1 &=&
	- \frac{2}{(D\!+\!1)} \Biggl[ 1 + \frac{2}{(D\!+\!1)}
		\frac{\partial}{\partial u} \Biggr]
		\Biggl\{ \frac{(\mu^2\!-\! \nu^2)}{4} (F_{\mu\nu})_2^0 + (Q_{\mu\nu}^-)_2 \Biggr\} \, ,
\\
(F_{\mu\nu})_2^2 &=&
	\frac{1}{(D\!+\!3)(D\!+\!2)} \Biggl[ 1 - \frac{2}{(D\!+\!2)} \frac{\partial}{\partial u} \Biggr]
		\Biggl\{
		\frac{(D\!+\!4) (\mu^2\!-\!\nu^2)}{4} (F_{\mu\nu})_2^1
\nonumber \\
&&	\hspace{-1.5cm}
		-  \Bigl( 1 \!-\! \frac{\partial}{\partial u} \Bigr)
			\biggl[ (D\!+\!1) \frac{\partial}{\partial u} + \frac{(D\!+\!1)^2}{4}
				- \frac{(\mu^2\!+\!\nu^2)}{2} \biggr] (F_{\mu\nu})_2^0
		\Biggr\}
	+ \frac{(Q_{\mu\nu}^+)_2}{(D\!+\!3)} \, . \qquad
\end{eqnarray}

\subsubsection{Final expressions}

Having solved for all the coefficients of the diagonal~$K$-propagators, we can
write the full solutions in a closed form,
\begin{eqnarray}
&&\hspace{-0.7cm}
K_{AA}(x;x') =
	\frac{e^{-u}}{H^2} \Biggl\{
	\frac{1}{4} I \bigl[ A(y) \bigr]
	+ \frac{H^{D-2} \, \Gamma\bigl( \frac{D-2}{2} \bigr)}{(4\pi)^{D/2}} 
		\times \frac{\Gamma(D\!-\!1)}{4 \, \Gamma\bigl( \frac{D}{2} \bigr) \, 
				\Gamma\bigl( \frac{D-2}{2} \bigr)} 
\nonumber \\
&&	\hspace{3.5cm}
\times
	\Biggl[
		- \frac{2 A_1  \Gamma\bigl( \frac{D}{2} \bigr)}{\Gamma(D\!-\!1)}
		- \frac{2(2D\!-\!3)}{(D\!-\!2)}
		- (2\!-\!y) u + y \Biggr]
	\Biggr\} \, ,
\\
&&\hspace{-0.7cm}
K_{BB}(x;x') =
	\frac{e^{-u}}{H^2} \Biggl\{
		\frac{1}{4} I \bigl[ B(y) \bigr]
		+ \frac{H^{D-2} \, \Gamma\bigl( \frac{D-2}{2} \bigr)}{(4\pi)^{D/2}} \Biggl[
			\frac{2}{(D\!-\!4)} - \frac{\Gamma(D\!-\!3)}{\Gamma^2\bigl( \frac{D-2}{2} \bigr)} 
				u
			\Biggr]
	\Biggr\} \, , \qquad
\\
&&\hspace{-0.7cm}
K_{CC}(x;x') =
	\frac{e^{-u}}{H^2} \Biggl\{
	\frac{1}{4} I \bigl[ C(y) \bigr] 
		+ \frac{H^{D-2} \, \Gamma\bigl( \frac{D-2}{2} \bigr)}{(4\pi)^{D/2}} 
			\Biggl[ \frac{ \Gamma(D\!-\!4)}
					{\Gamma^2\bigl( \frac{D-2}{2} \bigr) } 
						+ \frac{ e^{-(D-4)u} }{(D\!-\!4)} \Biggr]
	\Biggr\} \, , \qquad
\end{eqnarray}
where we define the primitive function~$I[f(y)] \!=\! \int^y \! dy' \, f(y')$.
Four relations deriving from the reflection identities~(\ref{reflection1}) 
and~(\ref{reflection2}),
\begin{eqnarray}
\partial_0 K_{AA} + \Bigl[ \partial_0' + (D\!-\!2) Ha' \Bigr] K_{BB} &=& 0 \, ,
\\
\partial_0' K_{AA} + \Bigl[ \partial_0 + (D\!-\!2) Ha \Bigr] K_{BB} &=& 0 \, ,
\\
\Bigl[ \partial_0 + Ha \Bigr] K_{BB} + \Bigl[ \partial_0' + (D\!-\!3) Ha' \Bigr] K_{CC} &=& 0 \, ,
\\
\Bigl[ \partial_0' + Ha' \Bigr] K_{BB} + \Bigl[ \partial_0 + (D\!-\!3) Ha \Bigr] K_{CC} &=& 0 \, ,
\end{eqnarray}
can be used to check the consistency of the solutions.

The off-diagonal~$K$-propagators can be given in terms of the power
series~(\ref{K ansatz}), where the coefficients of the non-integer powers
are given by~(\ref{E_0}-\ref{E_2}) and the recurrence relation~(\ref{D diff n}),
and the leading coefficients of the integer powers are 
given in~(\ref{leader}-\ref{last}), 
with Eqs.~(\ref{F diff 0}) and~(\ref{F sum n}) generating the higher coefficients.

The expressions for the~$J$-propagators are obtained 
from~(\ref{JAA}-\ref{JBC}), by acting
first-order the derivative operators on the~$K$-propagators whose solutions we have found
here. A number of relations that relate the~$J$-propagators with the~$I$-propagators
can be derived from reflection identities~(\ref{reflection1})
and~(\ref{reflection2}), {\it e.g.}
\begin{eqnarray}
\partial_0 J_{AA} + \Bigl[\partial_0' + (D\!-\!2) Ha'\Bigr] J_{BB}
& = & H I_{AB} \; , \label{JAABB1} \\
\partial_0' J_{AA} + \Bigl[\partial_0 + (D\!-\!2) Ha\Bigr] J_{BB}
& = & H I_{AB} \; . \label{JAABB2}
\end{eqnarray}
and can be used to independently check the validity of the solutions.


\begin{thebibliography}{99}

\bibitem{Tsamis:1996qk} 
  N.~C.~Tsamis and R.~P.~Woodard,
  Phys.\ Rev.\ D {\bf 54}, 2621 (1996)
  doi:10.1103/PhysRevD.54.2621
  [hep-ph/9602317].

\bibitem{Miao:2005am} 
  S.~P.~Miao and R.~P.~Woodard,
  Class.\ Quant.\ Grav.\  {\bf 23}, 1721 (2006)
  doi:10.1088/0264-9381/23/5/016
  [gr-qc/0511140].

\bibitem{Kahya:2007bc} 
  E.~O.~Kahya and R.~P.~Woodard,
  Phys.\ Rev.\ D {\bf 76}, 124005 (2007)
  doi:10.1103/PhysRevD.76.124005
  [arXiv:0709.0536 [gr-qc]].

\bibitem{Miao:2012bj} 
  S.~P.~Miao,
  Phys.\ Rev.\ D {\bf 86}, 104051 (2012)
  doi:10.1103/PhysRevD.86.104051
  [arXiv:1207.5241 [gr-qc]].

\bibitem{Leonard:2013xsa} 
  K.~E.~Leonard and R.~P.~Woodard,
  Class.\ Quant.\ Grav.\  {\bf 31}, 015010 (2014)
  doi:10.1088/0264-9381/31/1/015010
  [arXiv:1304.7265 [gr-qc]].

\bibitem{Mora:2013ypa} 
  P.~J.~Mora, N.~C.~Tsamis and R.~P.~Woodard,
  JCAP {\bf 1310}, 018 (2013)
  doi:10.1088/1475-7516/2013/10/018
  [arXiv:1307.1422 [gr-qc]].

\bibitem{Miao:2006gj} 
  S.~P.~Miao and R.~P.~Woodard,
  Phys.\ Rev.\ D {\bf 74}, 024021 (2006)
  doi:10.1103/PhysRevD.74.024021
  [gr-qc/0603135].

\bibitem{Kahya:2007cm} 
  E.~O.~Kahya and R.~P.~Woodard,
  Phys.\ Rev.\ D {\bf 77}, 084012 (2008)
  doi:10.1103/PhysRevD.77.084012
  [arXiv:0710.5282 [gr-qc]].

\bibitem{Wang:2014tza} 
  C.~L.~Wang and R.~P.~Woodard,
  Phys.\ Rev.\ D {\bf 91}, no. 12, 124054 (2015)
  doi:10.1103/PhysRevD.91.124054
  [arXiv:1408.1448 [gr-qc]].

\bibitem{Glavan:2013jca} 
  D.~Glavan, S.~P.~Miao, T.~Prokopec and R.~P.~Woodard,
  Class.\ Quant.\ Grav.\  {\bf 31}, 175002 (2014)
  doi:10.1088/0264-9381/31/17/175002
  [arXiv:1308.3453 [gr-qc]].

\bibitem{Tsamis:1992xa} 
  N.~C.~Tsamis and R.~P.~Woodard,
  Commun.\ Math.\ Phys.\  {\bf 162}, 217 (1994).
  doi:10.1007/BF02102015

\bibitem{Woodard:2004ut} 
  R.~P.~Woodard,
  gr-qc/0408002.

\bibitem{Mora:2012zi} 
  P.~J.~Mora, N.~C.~Tsamis and R.~P.~Woodard,
  J.\ Math.\ Phys.\  {\bf 53}, 122502 (2012)
  doi:10.1063/1.4764882
  [arXiv:1205.4468 [gr-qc]].
  
\bibitem{Glavan:2015ura} 
  D.~Glavan, S.~P.~Miao, T.~Prokopec and R.~P.~Woodard,
  Class.\ Quant.\ Grav.\  {\bf 32}, no. 19, 195014 (2015)
  doi:10.1088/0264-9381/32/19/195014
  [arXiv:1504.00894 [gr-qc]].

\bibitem{Glavan:2016bvp} 
  D.~Glavan, S.~P.~Miao, T.~Prokopec and R.~P.~Woodard,
  Class.\ Quant.\ Grav.\  {\bf 34}, no. 8, 085002 (2017)
  doi:10.1088/1361-6382/aa61da
  [arXiv:1609.00386 [gr-qc]].

\bibitem{Miao:2017feh}
  S.~P.~Miao, T.~Prokopec and R.~P.~Woodard,
  Phys.\ Rev.\ D {\bf 96}, no. 10, 104029 (2017)
  doi:10.1103/PhysRevD.96.104029
  [arXiv:1708.06239 [gr-qc]].
   
\bibitem{Donoghue:1993eb} 
  J.~F.~Donoghue,
  Phys.\ Rev.\ Lett.\  {\bf 72}, 2996 (1994)
  doi:10.1103/PhysRevLett.72.2996
  [gr-qc/9310024].

\bibitem{Donoghue:1994dn} 
  J.~F.~Donoghue,
  Phys.\ Rev.\ D {\bf 50}, 3874 (1994)
  doi:10.1103/PhysRevD.50.3874
  [gr-qc/9405057].

\bibitem{Donoghue:1996mt} 
  J.~F.~Donoghue and T.~Torma,
  Phys.\ Rev.\ D {\bf 54}, 4963 (1996)
  doi:10.1103/PhysRevD.54.4963
  [hep-th/9602121].
  
\bibitem{Miao:2009hb} 
  S.~P.~Miao, N.~C.~Tsamis and R.~P.~Woodard,
  J.\ Math.\ Phys.\  {\bf 50}, 122502 (2009)
  doi:10.1063/1.3266179
  [arXiv:0907.4930 [gr-qc]].

\bibitem{Capper:1979ej} 
  D.~M.~Capper,
  J.\ Phys.\ A {\bf 13}, 199 (1980).
  doi:10.1088/0305-4470/13/1/022

\bibitem{Leonard:2012fs} 
  K.~E.~Leonard and R.~P.~Woodard,
  Phys.\ Rev.\ D {\bf 85}, 104048 (2012)
  doi:10.1103/PhysRevD.85.104048
  [arXiv:1202.5800 [gr-qc]].
  
\bibitem{Onemli:2002hr} 
  V.~K.~Onemli and R.~P.~Woodard,
  Class.\ Quant.\ Grav.\  {\bf 19}, 4607 (2002)
  doi:10.1088/0264-9381/19/17/311
  [gr-qc/0204065].

\bibitem{Onemli:2004mb} 
  V.~K.~Onemli and R.~P.~Woodard,
  Phys.\ Rev.\ D {\bf 70}, 107301 (2004)
  doi:10.1103/PhysRevD.70.107301
  [gr-qc/0406098].

\bibitem{Miao:2017vly} 
  S.~P.~Miao, N.~C.~Tsamis and R.~P.~Woodard,
  Phys.\ Rev.\ D {\bf 95}, no. 12, 125008 (2017)
  doi:10.1103/PhysRevD.95.125008
  [arXiv:1702.05694 [gr-qc]].
  
\bibitem{Tsamis:1992zt} 
  N.~C.~Tsamis and R.~P.~Woodard,
  Phys.\ Lett.\ B {\bf 292}, 269 (1992).
  doi:10.1016/0370-2693(92)91174-8

\bibitem{Miao:2011fc} 
  S.~P.~Miao, N.~C.~Tsamis and R.~P.~Woodard,
  J.\ Math.\ Phys.\  {\bf 52}, 122301 (2011)
  doi:10.1063/1.3664760
  [arXiv:1106.0925 [gr-qc]].

\end{thebibliography}
\end{document}